\documentstyle[12pt]{article}
\begin{document}
\parskip=5pt plus 1pt minus 1pt

\begin{flushright}
{\large Ms $\#$388}\\
{\large Revised}
\end{flushright}

\vspace{0.2cm}

\begin{center}
{\Large\bf A Further Study of $CP$ Asymmetries}\\
{\Large\bf in Pure Penguin-induced $B^{\pm}_{u}$ Decays}
\end{center}

\vspace{0.4cm}

\begin{center}
{\bf Dongsheng DU}\footnote{Supported in part by the National Nature Science
Foundation of China} \\
{\sl Institute of High Energy Physics, P.O. Box 918(4), Beijing 100039, P.R.
China}
\end{center}

\begin{center}
and
\end{center}

\begin{center}
{\bf Zhi-zhong XING}\footnote{Alexander-von-Humboldt Research Fellow} \\
{\sl Sektion Physik der Universit$\ddot{\rm a}$t M$\ddot{\rm u}$nchen, D-80333
Munich, Germany}
\end{center}

\vspace{0.6cm}

\begin{abstract}

	We make a further study of $CP$ asymmetries in the pure penguin-induced decay
modes
$B^{-}_{u}\rightarrow K^{(*)0} +  K^{(*)-}$ and $B^{-}_{u}
\rightarrow \bar{K}^{(*)0} + (\pi^{-}, \rho^{-}, a^{-}_{1})$
by using the two-loop renormalization-group-improved
effective Hamiltonian and factorization approximation.
Different from the previous results obtained without QCD corrections,
the $CP$ asymmetries in each of the two
groups are enhanced and classified with their respective
factorization coefficients.
A sum over the pure penguin modes of each group is proposed to obtain an
effective branching ratio and a {\it weighted-mean} signal of
$CP$ violation, which should be statistically significant for experimental
observation.

\end{abstract}

\newpage

\begin{flushleft}
{\large\bf 1 $~$ Introduction}
\end{flushleft}

	Observing $CP$ violation in the $B$-meson system and
confronting it with the predictions of the standard model is an important
task in particle physics. The basic signal for $CP$
violation is the partial rate difference between
a $B$ decay mode and its $CP$-conjugate process. On the basis of the
Cabibbo-Kobayashi-Maskawa (CKM) picture, either $B^{0}-\bar{B}^{0}$
mixing [1] or the absorptive parts of QCD-loop-induced (penguin) amplitudes
[2] can give rise to significant $CP$ asymmetries in exclusive
nonleptonic $B$ decays. Among various hadronic decay modes of $B$ mesons,
the pure penguin-induced transitions
are of great interest to explore direct $CP$ violation in the decay
amplitude and to test our understanding of the loop effects involved in
the quark-level process $b\rightarrow q $ (with $q=d$ or $s$).
Recently evidence for the electromagnetic penguin transitions, e.g.,
$B^{0}_{d}\rightarrow K^{*0}\gamma$ and $B^{-}_{u}\rightarrow K^{*-}
\gamma$, has been obtained by CLEO Collaboration [3]. Accordingly studying
$CP$ violation in the pure strong penguin-induced $B$
decays becomes a necessary and realistic topic today.

	In the literature [4-6], $CP$ asymmetries in some pure penguin channels have
been calculated
with the QCD-uncorrected effective Hamiltonian $H^{\rm pen}_{\rm eff}$ (see
Eq.(2))
and factorization approximation. For the processes occurring through
$b\rightarrow d$ (or
$b\rightarrow s$), those results are approximately independent of
the spin properties of final-state mesons.
Hence a sum over the relevant modes has been proposed in order to reduce the
total number of $B$ events
and to obtain a statistically significant signal of $CP$ violation.
In this paper, we
shall use the two-loop renormalization-group-improved effective Hamiltonian
${\cal H}_{\rm eff}$ (see Eq.(4)) to make a further study of $CP$ asymmetries
in
the following two handfuls of pure penguin transitions:
\begin{equation}
\begin{array}{llll}
{\rm (I)} & B^{-}_{u} & \longrightarrow & K^{0}K^{-},\;
K^{*0}K^{-},\; K^{0}K^{*-},\; K^{*0}K^{*-}\; ; \\
{\rm (II)} & B^{-}_{u} & \longrightarrow & \bar{K}^{0}\pi^{-},\;
\bar{K}^{0}\rho^{-},\; \bar{K}^{0}a_{1}^{-},\; \bar{K}^{*0}\pi^{-},\;
\bar{K}^{*0}\rho^{-},\; \bar{K}^{*0}a_{1}^{-}\; .
\end{array}
\end{equation}
It is worth emphasizing that the above ten decay modes are flavor
self-tagging processes, which should be favored for experimental
reconstructions.
In addition, there are no QCD-loop-induced hairpin
diagrams contributing to these transitions\footnote{In our sense, some of the
QCD-loop-induced transitions
such as $B^{-}_{u}\rightarrow K^{-}\phi$ are not the pure penguin decay modes.
The reason is that they can also occur through the hairpin diagrams. For a
detailed discussion, see Ref. [7].}
and the electroweak penguin effects on them are negligible [8]. Thus
application of the factorization approximation
to these decays should lead to less uncertainty.
Different from the naive results obtained with $H^{\rm pen}_{\rm eff}$,
the magnitude of $CP$ asymmetries in the transitions (I) and (II) are enhanced
by the next-to-leading order QCD corrections from ${\cal H}_{\rm eff}$.
In each group the asymmetries become
non-degenerate and are classified with their
respective factorization coefficients, but they still keep the same sign.
Hence
we propose a sum over the modes of each group to obtain an effective
branching ratio and a {\it weighted-mean} signal of
$CP$ violation, which should be
statistically significant for observation. The feasibility of measuring such
weighted-mean $CP$ asymmetries the uncertainties in evaluating them are briefly
discussed. \\

\begin{flushleft}
{\large\bf 2 $~$ Effective Hamiltonians and factorization approximation}
\end{flushleft}

	To calculate decay amplitudes of the aforelisted pure
penguin channels, we use the low-energy effective Hamiltonians for
$\Delta B=\pm 1$ transitions and the well-known factorization approximation
[9].
Neglecting influence of the renormalization group, one used to apply
the one-loop penguin Hamiltonian [4-6]
\begin{equation}
H^{\rm pen}_{\rm eff}(\Delta B=-1)\;=\; -\frac{G_{F}}{\sqrt{2}}
\frac{\alpha_{s}(m_{b})}{8\pi} \left [\sum_{i=u,c,t}
v_{i}F_{i}(k^{2})\right ]
\left (-\frac{Q_{3}}{N_{c}}+Q_{4}-\frac{Q_{5}}{N_{c}}+Q_{6}\right )\;
\end{equation}
to phenomenological discussions. In $H^{\rm pen}_{\rm eff}$,
the QCD renormalization effect is assumed to be approximately included by the
effective coupling constant $\alpha_{s}$ at the physical scale
$\mu =m_{b}$; $v_{i}\equiv V_{ib}V^{*}_{iq}$ (with $i=u,c,t$
and $q=d,s$) are the CKM factors corresponding to
$b\rightarrow q$; $F_{i}(k^{2})$ are the loop integral functions of
the time-like penguin diagrams versus the virtual gluon momentum
$k^{2}$ [2,6]; $N_{c}$ is the number of colors; and $Q_{3,...,6}$
represent the penguin operators. Together with the current-current operators
$Q^{u,c}_{1,2}$, $Q_{3,...,6}$ form an operator basis as follows:
\begin{eqnarray}
& Q^{u}_{1} & =\; (\bar{q}_{\alpha}u_{\beta})^{~}_{V-A}
(\bar{u}_{\beta}b_{\alpha})^{~}_{V-A}\; ,\;\;\;\;\;
Q^{u}_{2}\; =\; (\bar{q}u)^{~}_{V-A}(\bar{u}b)^{~}_{V-A}\; ,\nonumber\\
& Q^{c}_{1} & =\; (\bar{q}_{\alpha}c_{\beta})^{~}_{V-A}
(\bar{c}_{\beta}b_{\alpha})^{~}_{V-A}\; ,\;\;\;\;\;
Q^{c}_{2}\; =\; (\bar{q}c)^{~}_{V-A}(\bar{c}b)^{~}_{V-A}\; ,\nonumber\\
& Q_{3} & =\; (\bar{q}b)^{~}_{V-A}
\sum_{q^{'}}(\bar{q}^{'}q^{'})^{~}_{V-A}\; ,\;\;\;\;\;
Q_{4}\; =\; (\bar{q}_{\alpha}b_{\beta})^{~}_{V-A}
\sum_{q^{'}}(\bar{q}^{'}_{\beta}q^{'}_{\alpha})^{~}_{V-A}\; ,\\
& Q_{5} & =\; (\bar{q}b)^{~}_{V-A}
\sum_{q^{'}}(\bar{q}^{'}q^{'})^{~}_{V+A}\; ,\;\;\;\;\;
Q_{6}\; =\; (\bar{q}_{\alpha}b_{\beta})^{~}_{V-A}
\sum_{q^{'}}(\bar{q}^{'}_{\beta}q^{'}_{\alpha})^{~}_{V+A}\; .\nonumber
\end{eqnarray}
Compared with $H^{\rm pen}_{\rm eff}$, the two-loop QCD-corrected Hamiltonian
${\cal H}_{\rm eff}$ is of the form [10]
\begin{equation}
{\cal H}_{\rm eff}(\Delta B=-1)\;=\;\frac{G_{F}}{\sqrt{2}}
\left [v_{u}\left (\sum_{i=1}^{2}c_{i}Q^{u}_{i}\right )
+v_{c}\left (\sum_{i=1}^{2}c_{i}Q_{i}^{c}\right )
-v_{t}\left (\sum_{i=3}^{6}c_{i}Q_{i}\right )\right ]\; .
\end{equation}
Here the Wilson coefficient functions
$c_{1,...,6}$, obtained by applying the renormalization-group-improved
perturbation theory,
have included the next-to-leading order QCD corrections at the scale
$\mu =m_{b}$. Note that $c_{i}$ depend on the renormalization
scheme used for the four-quark operators $Q_{i}$. This scheme dependence,
however,
can be cancelled by certain one-loop matrix elements of the decay mode in
question.
For a detailed description of the approach to
obtain the renormalization-scheme independent transition amplitudes in two-body
$B$ decays,
we refer the reader to the article of Fleischer [11].

	With the help of ${\cal H}_{\rm eff}$, the amplitude of
a two-body pure penguin-induced decay mode $B^{-}_{u}\rightarrow XY$
may be expressed as linear combinations of $<XY|Q_{i}|B^{-}_{u}>$
with weighting factors $c_{i}$ and $v_{i}$. Following Ref. [11], one can remove
the
renormalization-scheme dependence of $c_{i}$ in the physical transition
amplitudes by taking one-loop penguin matrix elements of the operators
$Q^{u,c}_{2}$ into account (see Fig. 1).
The renormalization-scheme independent amplitude of $B^{-}_{u}\rightarrow XY$
is finally obtained as
[11,12]
\begin{eqnarray}
&   & <XY|{\cal H}_{\rm eff}(\Delta B=-1)|B^{-}_{u}>
\nonumber \\
& = & -\frac{G_{F}}{\sqrt{2}}\left \{ \bar{c}_{2}\left [
v_{u}F_{u}(k^{2})+v_{c}F_{c}(k^{2})+v_{t}\left (\frac{10}{9}
-\frac{2}{3}\ln \frac{m^{2}_{b}}{m^{2}_{W}}
\right )\right ]M_{P}+v_{t}M'_{P}\right \}\; ,
\end{eqnarray}
where $\bar{c}_i$ (with $i=2,...,6$) are the scheme independent Wilson
coefficients
including the next-to-leading order QCD corrections [10], and $M_{P}$ and
$M^{'}_{P}$
are the hadronic matrix
elements of penguin operators defined by
\begin{eqnarray}
M_{P} & \equiv & \frac{\alpha_{s}(m_{b})}{8\pi}
<XY|\left (-\frac{Q_{3}}{N_{c}}+Q_{4}-\frac{Q_{5}}{N_{c}}+Q_{6}\right )
|B^{-}_{u}> \; ,\nonumber\\
M^{'}_{P} & \equiv & \sum^{6}_{i=3}\left (\bar{c}_{i}<XY|Q_{i}|B^{-}_{u}>
\right )\; .
\end{eqnarray}
In contrast, the decay amplitude of $B^{-}_{u}\rightarrow XY$ calculated with
$H^{\rm pen}_{\rm eff}$
is of the form
\begin{equation}
<XY|H^{\rm pen}_{\rm eff}(\Delta B=-1)|B^{-}_{u}>\; =\;
-\frac{G_{F}}{\sqrt{2}}\left [\sum_{i=u,c,t}v_{i}F_{i}(k^{2})\right ] M_{P} \;
{}.
\end{equation}

	The hadronic matrix elements $M_{P}$ and $M^{'}_{P}$ are calculated using the
naive
factorization approximation [9]. Here we only consider contributions from
the spectator-type diagrams as illustrated in Fig. 1, while the OZI-forbidden
transitions and (or) formfactor-suppressed annihilation topologies are
neglected.
Consequently $M_{P}$ and $M^{'}_{P}$ are both dominated by a
single term
\begin{equation}
M^{XY}_{qq'q'} \equiv <X(q\bar{q}')|(\bar{q}q')^{~}_{V-A}|0>
<Y(q'\bar{u})|(\bar{q}'b)^{~}_{V-A}|B^{-}_{u}(b\bar{u})> \; ,
\end{equation}
and their factorized expressions are given as
\begin{eqnarray}
M_{P} & = & \left [\frac{\alpha_{s}(m_{b})}{8\pi}\left (
1-\frac{1}{N_{c}^{2}}\right )(1+\zeta_{P})\right ]M^{XY}_{qq'q'}\; ,
\nonumber \\
M'_{P} & = & \left [\left (\frac{\bar{c}_{3}}{N_{c}}+\bar{c}_{4}
\right )+\left (\frac{\bar{c}_{5}}{N_{c}}+\bar{c}_{6}\right )
\zeta_{P}\right ]M^{XY}_{qq'q'}\; .
\end{eqnarray}
In Eq. (9), the factorization coefficient $\zeta_{P}$ is obtained by
transforming
the $(V-A)(V+A)$ (or $(S+P)(S-P)$) currents of $Q_{5,6}$ into the
$(V-A)(V-A)$ ones. For each specific decay mode, $\zeta_{P}$ depends upon the
current masses of quarks as well as
the angular momenta and parities of $X$ and $Y$ mesons. We list
the expressions of $\zeta_{P}$ in Table 1 for $0^{\pm}0^{\pm},
0^{\pm}1^{\pm}, 1^{\pm}0^{\pm}$, and $1^{\pm}1^{\pm}$ final states,
respectively.
The hadronic matrix elements $M^{XY}_{qq'q'}$
can be Lorentz-invariantly decomposed with the relevant decay constants
and formfactors, as given in Refs. [6,12]. For our purpose, we shall sum over
the
polarizations of vector or axial vector mesons in the final state and estimate
$|M^{XY}_{qq'q'}|^{2}$ by means of the Bauer-Stech-Wirbel (BSW) model [13]. \\

\begin{flushleft}
{\large\bf 3 $~$ $CP$ asymmetries}
\end{flushleft}

	Now let us define the $CP$-violating partial decay rate
asymmetry for $B^{-}_{u}\rightarrow f$ and its $CP$-conjugate counterpart
$B^{+}_{u}\rightarrow \bar{f}$:
\begin{equation}
{\cal A}_{CP}(f)\equiv \frac{\Gamma (B^{+}_{u}\rightarrow \bar{f})
-\Gamma (B^{-}_{u}\rightarrow f)}
{\Gamma (B^{+}_{u}\rightarrow \bar{f})
+\Gamma (B^{-}_{u}\rightarrow f)} \; .
\end{equation}
In the CKM mechanism with three families of quarks, every of the aforementioned
pure penguin transitions contains two amplitude components
that have different
$CP$-violating weak phases ($v_{u}$ and $v_{c}$) and different
$CP$-conserving strong phases (${\rm Im}F_{u}(k^{2})$ and ${\rm
Im}F_{c}(k^{2})$).
The $CP$ asymmetry is approximately independent of hadronic matrix elements,
since $M^{XY}_{qq'q'}$ can be cancelled in ${\cal A}_{CP}(f)$.
Thus ${\cal A}_{CP}(f)$ is simply expressed as [6]
\begin{equation}
{\cal A}_{CP}(f)\;=\; \frac{-2{\rm Im}(v_{u}v^{*}_{c})\cdot {\rm Im}R_{f}}
{|v_{u}|^{2}+|v_{c}|^{2}\cdot |R_{f}|^{2}+2{\rm Re}(v_{u}v^{*}_{c})
\cdot {\rm Re}R_{f}}\; ,
\end{equation}
where $R_{f}$ is a ratio of strong-interaction parts of the
two amplitude components corresponding to $v_{c}$ and $v_{u}$.
When calculating ${\cal A}_{CP}(f)$ with $H^{\rm pen}_{\rm eff}$, we obtain
[4-6]
\begin{equation}
R_{f}\;=\; \frac{F_{c}(k^{2})-F_{t}(k^{2})}
{F_{u}(k^{2})-F_{t}(k^{2})}\; .
\end{equation}
This implies that $CP$ asymmetries in the transitions (I) or (II)
will have the same magnitude and the same sign.
While applying ${\cal H}_{\rm eff}$ to the analysis,
$R_{f}$ becomes
\begin{equation}
R^{'}_{f}\; =\; \frac{F_{c}(k^{2})-S(\zeta_{P})}{F_{u}(k^{2})-S(\zeta_{P})}\; ,
\end{equation}
where
\begin{equation}
\displaystyle
S(\zeta_{P})\; = \; \frac{10}{9}-\frac{2}{3}\ln \frac{m^{2}_{b}}{m^{2}_{W}}
+\frac{8\pi}{\bar{c}_{2}\cdot \alpha_{s}(m_{b})}\cdot
\frac{\left (\displaystyle\frac{\bar{c}_{3}}{N_{c}}+c_{4}\right )+
\left (\displaystyle\frac{\bar{c}_{5}}{N_{c}}+c_{6}\right )\zeta_{P}}
{\left (1-\displaystyle\frac{1}{N^{2}_{c}}\right )(1+\zeta_{P})}\; .
\end{equation}
Clearly $R^{'}_{f}$ includes the next-to-leading order QCD corrections and
depends upon the factorization parameter $\zeta_{P}$.
Since the values of $\zeta_{P}$ for
$B^{-}_{u}\rightarrow K^{0}K^{-}, K^{0}K^{*-}$, and $K^{*0}K^{-}$
(or $K^{*0}K^{*-}$) are different from one another,
their $CP$ asymmetries
will become non-degenerate in contrast with the result obtained from Eq. (12).
So is the situation for the pure penguin channels
$B^{-}_{u}\rightarrow \bar{K}^{(*)0} + (\pi^{-}, \rho^{-}, a_{1}^{-})$.
This feature, arising from the QCD corrections, requirs us to reconsider the
idea of suming over the pure penguin transitions (I) or (II)
to gain a statistically significant $CP$-violating signal.

	A comparison between $R^{'}_{f}$ and $R_{f}$ shows that
the QCD improvement only modifies
${\rm Re}F_{u,c}(k^{2})$. Hence the asymmetries in each group of decay modes
remain the same sign. This ensures that there should be little dilution effect
on the
$CP$ asymmetries if one sums over the
available final states $f$ and $\bar{f}$ in the transitions (I) or (II).
In this case, one can obtain an effective branching ratio like
\begin{equation}
B_{\rm eff} \; \equiv \; \displaystyle\sum_{f}B(f) \; = \; \tau^{~}_{B}
\displaystyle
\sum_{f}\Gamma (B^{-}_{u}\rightarrow f) \; ,
\end{equation}
where $\tau^{~}_{B}$ is the lifetime of $B^{-}_{u}$ meson. The corresponding
$CP$ asymmetry ${\cal A}_{CP}$
is a {\it weighted-mean} value of ${\cal A}_{CP}(f)$:
\begin{equation}
{\cal A}_{CP}\; \equiv \; \frac{\left [\displaystyle\sum_{f}\Gamma
(B^{+}_{u}\rightarrow \bar{f})\right ]-\left [\displaystyle\sum_{f}\Gamma
(B^{-}_{u}\rightarrow f)\right ]}
{\left [\displaystyle\sum_{f}\Gamma (B^{+}_{u}\rightarrow \bar{f})\right ]
+\left [\displaystyle\sum_{f}\Gamma (B^{-}_{u}\rightarrow f)\right ]}
\; = \; \frac{\displaystyle\sum_{f}\left \{ {\cal A}_{CP}(f)\left [
B(\bar{f})+B(f) \right ] \right \}}
{\displaystyle\sum_{f}\left [B(\bar{f}) +B(f)\right ]} \; .
\end{equation}
Although the value of ${\cal A}_{CP}$ is impossible to deviate too much from
the ones of ${\cal A}_{CP}(f)$, it is expected that the effective branching
ratio
$B_{\rm eff}$ can become several times larger than a single $B(f)$.
Thus the total number of $B^{\pm}_{u}$ events needed in measuring
${\cal A}_{CP}$ will be remarkably reduced and could be accessible
in the forthcoming $B$-meson factories.
Note that the weighted-mean
asymmetry depends upon hadronic matrix elements and is therefore difficult to
be evaluated in a reliable way. In the long run, however, a great improvement
of the present calculations will be  possible  to yield trustworthy results. \\

\begin{flushleft}
{\large\bf IV $~$ Numerical results and discussion}
\end{flushleft}

	For illustration,
we make a numerical estimate of $CP$ asymmetries and branching ratios
for the pure penguin-induced decay modes (I) and (II). The CKM parameters
are taken as $\lambda =0.22$, $A=0.90$, $\rho=-0.50$, and $\eta=0.30$,
which are consistent with the present data on $B^{0}_{d}-\bar{B}^{0}_{d}$
mixing and $\epsilon^{~}_{K}$ [14]. We choose the values of the current
quark masses as $m_{u}=5$ MeV, $m_{d}=10$ MeV, $m_{s}=175$ MeV, $m_{c}
=1.35$ GeV, $m_{b}=4.8$ GeV, and $m_{t}=150$ GeV. Fixing $\Lambda
^{(5)}_{\overline{MS}}=200$ MeV, the effective QCD coupling constant
and next-to-leading order Wilson coefficients are obtained [10] as $\alpha_{s}
(m_{b})=0.18$, $\bar{c}_{1}=-0.273$, $\bar{c}_{2}=1.123$, $\bar{c}_{3}=0.014$,
$\bar{c}_{4}=-0.032$, $\bar{c}_{5}=0.009$, and $\bar{c}_{6}=-0.039$.
With the approximate rule of discarding $1/N_{c}$ corrections in the naive
factorization approach [9], here we take $1/N_{c}=0$ for the relevant
charmless nonleptonic $B$ transitions. It is not clear what value of $k^{2}$
(the virtual gluon momentum in penguin graphs) should be
taken in exclusive $B$ decays. A simple estimate, involving two-body
kinematics and one-gluon exchange to accelerate the spectator
quark, yields $m^{2}_{b}/4 \leq k^{2} \leq m^{2}_{b}/2$ [4].
To evaluate branching ratios we use $\tau^{~}_{B}=1.29\times 10^{-12}$
s [15] and quote values of the relevant decay constants and formfactors
from Ref. [13].

	Our numerical results of $CP$ asymmetries ${\cal A}_{CP}(f)$
and branching ratios $B(f)$ are given in Fig. 2 and Tables 2 and 3,
respectively, as functions of $k^{2}$. The following features can be
observed:

	(1) Compared with the previous result without the next-to-leading order
QCD corrections, now $CP$ asymmetries in the transitions (I) or (II)
are enhanced to some extent
and become non-degenerate for the $0^{\pm}0^{\pm}$, $0^{\pm}1^{\pm}$, and
$1^{\pm}0^{\pm}$ (or $1^{\pm}1^{\pm}$) final states.
As expected, ${\cal A}_{CP}(f)$ keep the same sign even though their values
may change significantly with $k^{2}$.

	(2) Among the transitions (I), the $0^{\pm}1^{\pm}$-type decay
mode $B^{-}_{u}\rightarrow K^{0}K^{*-}$ has the
largest $CP$ asymmetry but the smallest branching ratio. In contrast,
the $0^{\pm}0^{\pm}$-type channel $B^{-}_{u}\rightarrow K^{0}K^{-}$
has a relatively small $CP$ asymmetry in spite of its largest branching ratio.
So is the situation for the $0^{\pm}1^{\pm}$- and $0^{\pm}0^{\pm}$-type
decay modes in the transitions (II). One can find that there exists no
individual channel which has obvious advantages over
the others of the group for measuring $CP$ violation.
For this reason, studying the weighted-mean $CP$ asymmetries should be
interesting in statistics.

	(3) For either the transitions (I) or (II), the weighted-mean
asymmetry ${\cal A}_{CP}$ is almost of the same magnitude as each ${\cal
A}_{CP}(f)$,
but its corresponding effective branching ratio $B_{\rm eff}$
is several times larger than the individual $B(f)$. For example,
$B_{\rm eff}/B(K^{0}K^{-}) \sim 2.5$,
$B_{\rm eff}/B(K^{0}K^{*-}) \sim 7$,
$B_{\rm eff}/B(K^{*0}K^{-}) \sim 4$, and
$B_{\rm eff}/B(K^{*0}K^{*-}) \sim 3.5$.
As a result, measuring ${\cal A}_{CP}$ instead of ${\cal A}_{CP}(f)$
should reduce the needed $B^{\pm}_{u}$ events several times and
is therefore favored in practice. At the $3\sigma$ level, our estimates
indicate that about $5\times 10^{8}$ and $2\times 10^{9}$
$B^{\pm}_{u}$ pairs are available to probe the weighted-mean $CP$ asymmetries
of
$7\%$ and $1\%$, respectively, in the transitions (I) and (II).
In experiments, to detect a handful of pure penguin channels with similar
final-state particles (e.g., $B^{-}_{u}\rightarrow K^{(*)0} + K^{(*)-}$)
might not be more difficult than to detect an individual channel among them.

	Except for the uncertainty induced by the CKM parameters, there are
some other theoretical uncertainties from the assumptions and approximations
made in our calculations. It is worthwhile at this point to give a brief
summary of them:

	(1) For the pure penguin-induced exclusive $B$ decays in question, $CP$
violation has been treated by postulating that the phases of the penguin
amplitudes
at the meson level is the same as those of the penguin loops at the quark
level.
This leads to the problem that the $CP$ asymmetries ${\cal A}_{CP}(f)$ depend
strongly upon the virtual gluon momentum ($k^{2}$) of the timelike penguin
graphs. The large uncertainty arised from $k^{2}$ cannot be experimentally
limited in a meaningful way, since it is not a measurable. To modify this
unsatisfactory point, an attempt has been made by calculating the $k^{2}$
distribution of some exclusive $B$ decays and
fold it with the momentum dependence of the loop amplitudes [16].

	(2) Neglecting inelastic rescattering effects, pure penguin decay modes have
only one isospin component. Hence their $CP$ asymmetries should suffer little
from
the unknown final-state interactions. Under the same condition, a sum over the
transitions (I) or (II) is in principle available without suffering
cancellations.
Whether the above simplification is reasonable or not can be clarified in the
forthcoming experiments of $B$-meson physics.

	(3) It is seen that the $CP$ asymmetries ${\cal A}_{CP}(f)$ are independent
of hadronic matrix elements in the factorization approximation. However, the
branching ratios $B(f)$ should be sensitive to the model applied to hadronic
matrix elements. Although the weighted-mean asymmetries ${\cal A}_{CP}$ are
also
affected by hadronic physics, the involved uncertainty should not be drastic in
general. The reason is that both ${\cal A}_{CP}(f)$ and ${\cal A}_{CP}$ are
ratios of the partial decay rates. As an illustration, our
numerical estimates of ${\cal A}_{CP}$ and $B_{\rm eff}$ could give one a
feeling
of ballpark numbers to be expected.

	In conclusion, the pure penguin-induced $B^{\pm}_{u}$ decays
are of great interest to uncover direct $CP$ violation in the
decay amplitude. A weighted-mean signal of $CP$ asymmetries may be
more easily observed in experiments, although to evaluate it
with reliability remains difficult in theories.
In order to test the strong-interaction penguin picture and direct
$CP$ violation mechanism at $B$-meson factories,
a further study of the dynamics of nonleptonic exclusive $B$ decays
is urgent. \\

\begin{flushleft}
{\large\bf Acknowledgements}
\end{flushleft}

	This paper was revised when one of the authors (Z.Z.X.) visited
Z$\rm\ddot u$rich Universit$\rm\ddot a$t. He would like to thank Professor
D. Wyler for his hospitality and encouragement.

\newpage

\begin{flushleft}
{\Large\bf Figure Captions}
\end{flushleft}

\vspace{0.4cm}

{\bf Fig. 1} $~$ Pure penguin-induced decay modes $B^{-}_{u}\rightarrow
X(q\bar{q}') + Y(q'\bar{u})$ with $(q, q')=(d, s)$ or $(s, d)$ :
(a) the tree-level matrix elements of the penguin operators, and
(b) the one-loop penguin matrix elements of the current-current operators. \\

{\bf Fig. 2} $~$ $CP$ asymmetries versus $k^{2}$ (virtual gluon
momentum in penguin graphs) for a few pure penguin-induced $B^{\pm}_{u}$
decay modes: (a) exclusive $b\rightarrow d$, and (b) exclusive
$b\rightarrow s$. The circled curve is the degenerate
result without QCD corrections;
the dashed, dotted, and dot-dashed curves
stand for the QCD-corrected results;
and the dagged curve is the
weighted-mean $CP$-violating signal.

\newpage

\setcounter{page}{12}

\begin{flushleft}
{\Large\bf Table Captions}
\end{flushleft}

\vspace{0.4cm}

{\bf Table 1} $~$  Factorization parameters $\zeta_{P}$ (defined
in Eq. (9)) for the decay modes $B^{-}_{u}\rightarrow XY$, where $X$ and $Y$
may be $0^{\pm}$ or $1^{\pm}$ mesons. \\

{\bf Table 2} $~$ Branching ratios $B(f)$ ($\times 10^{-7}$)
versus $k^{2}$ (virtual gluon momentum in penguin graphs)
for $B^{-}_{u}\rightarrow K^{(*)0} + K^{(*)-}$. $B_{\rm eff}$
($\times 10^{-7}$) is an effective branching ratio obtained by summing
over these four pure penguin modes. \\

{\bf Table 3} $~$ Branching ratios $B(f)$ ($\times 10^{-6}$)
versus $k^{2}$ (virtual gluon momentum in penguin graphs)
for $B^{-}_{u}\rightarrow \bar{K}^{(*)0} + (\pi^{-}, \rho^{-}, a_{1}^{-})$.
$B_{\rm eff}$ ($\times 10^{-6}$) is an effective branching
ratio obtained by summing over these six pure penguin modes. \\

\vspace{2.5cm}

\begin{center}
{\bf Table 1}
\end{center}

\begin{center}
\begin{tabular}{c|c} \hline\hline
$(J^{P}_{X},\; J^{P}_{Y})\;\;\;\;$ 	& $\;\;\;\; \zeta_{P}$  \\ \hline \\
$(0^{\pm},\; 0^{\pm})$		& $\displaystyle \frac{2m^{2}_{X}}
{(m_{q}+m_{q'})m_{b}}$ \\ \\
$(0^{\pm},\; 1^{\pm})$		& $\displaystyle -\frac{2m^{2}_{X}}
{(m_{q}+m_{q'})m_{b}}$ \\ \\
$(1^{\pm},\; 0^{\pm})$		& 0 \\ \\
$(1^{\pm},\; 1^{\pm})$		& 0 \\ \\ \hline\hline
\end{tabular}
\end{center}

\newpage

\begin{center}
{\bf Table 2}
\end{center}

\begin{center}
\begin{tabular}{c|cccc|c} \hline\hline
$k^{2}/m^{2}_{b}$ & $B(K^{0}K^{-})$	& $B(K^{0}K^{*-})$
& $B(K^{*0}K^{-})$	& $B(K^{*0}K^{*-})$
& $B_{\rm eff}$ \\ \hline
0.1	& 12.5	& 0.39	& 7.47  & 9.20	& 29.6 \\
0.2	& 12.5	& 0.39  & 7.47	& 9.20	& 29.6 \\
0.3	& 14.2	& 0.46  & 8.54	& 10.5	& 33.7 \\
0.4	& 15.8	& 0.53  & 9.56	& 11.8	& 37.7 \\
0.5	& 14.6	& 0.49  & 8.86	& 10.9	& 34.8 \\
0.6	& 13.8	& 0.46  & 8.32	& 10.2	& 32.8 \\
0.7	& 13.1	& 0.43  & 7.87	& 9.69	& 31.1 \\
0.8	& 12.5	& 0.41  & 7.53	& 9.27	& 29.7 \\
0.9	& 12.0	& 0.39  & 7.22	& 8.89	& 28.5 \\
1.0	& 11.6	& 0.38  & 6.98	& 8.59	& 27.6 \\ \hline\hline
\end{tabular}
\end{center}

\vspace{1.5cm}

\begin{center}
{\bf Table 3}
\end{center}

\begin{center}
\begin{tabular}{c|cccccc|c} \hline\hline
$k^{2}/m^{2}_{b}$ & $B(\bar{K}^{0}\pi^{-})$	& $B(\bar{K}^{0}\rho^{-})$
& $B(\bar{K}^{0}a_{1}^{-})$	& $B(\bar{K}^{*0}\pi^{-})$
& $B(\bar{K}^{*0}\rho^{-})$	& $B(\bar{K}^{*0}a_{1}^{-})$
& $B_{\rm eff}$ \\ \hline
0.1	& 7.77	& 0.22 	& 0.19	& 4.84	& 5.99	& 5.39	& 24.4 \\
0.2	& 8.77	& 0.25	& 0.23	& 5.50	& 6.81	& 6.13	& 27.7 \\
0.3	& 11.3	& 0.36	& 0.32	& 7.20	& 8.92	& 8.03	& 36.1 \\
0.4	& 12.0	& 0.39	& 0.35	& 7.67	& 9.50	& 8.55	& 38.5 \\
0.5	& 10.7	& 0.34	& 0.30	& 0.81	& 8.43	& 7.59	& 34.2 \\
0.6	& 9.79	& 0.31	& 0.28	& 6.23	& 7.71	& 6.94	& 31.3 \\
0.7	& 9.13	& 0.28  & 0.25	& 5.79	& 7.17	& 6.45	& 29.1 \\
0.8	& 8.65	& 0.27	& 0.24	& 5.47	& 6.78	& 6.10	& 27.5 \\
0.9	& 8.22	& 0.25	& 0.22	& 5.19	& 6.43	& 5.79	& 26.1 \\
1.0	& 7.90	& 0.24	& 0.21	& 4.99	& 6.17	& 5.55	& 25.1 \\ \hline \hline
\end{tabular}
\end{center}

\end{document}